\documentstyle[amstex,amssymb,12pt,epsfig]{article}

\begin{document}

\begin{center}
{\large NEW RESULTS ON THE PHASE DIAGRAM OF THE}

{\large FFXY MODEL: A TWISTED CFT APPROACH}

\bigskip \bigskip

G. CRISTOFANO and V. MAROTTA

{\footnotesize Dipartimento di Scienze Fisiche and INFN, Universit\`{a} di
Napoli ``Federico II'',}

{\footnotesize Via Cintia, Compl. Universitario M. Sant'Angelo, 80126
Napoli, Italy}

{\footnotesize E-mail: cristofano@@na.infn.it}

\bigskip 

P. MINNHAGEN

{\footnotesize Department of Theoretical Physics, Umea University,}

{\footnotesize 901 87 Umea, Sweden}

{\footnotesize E-mail: Petter.Minnhagen@@physics.umu.se}

\bigskip 

A. NADDEO

{\footnotesize CNISM, Unit%
$\mu$%
a di Ricerca di Salerno,}

{\footnotesize Via Salvador Allende, 84081 Baronissi (SA), Italy}

{\footnotesize E-mail: naddeo@@sa.infn.it}

\bigskip 

G. NICCOLI

{\footnotesize LPTM, Universit\'{e} de Cergy-Pontoise,}

{\footnotesize 2 avenue Adolphe Chauvin, 95302 Cergy-Pontoise, France}

{\footnotesize E-mail: Giuliano.Niccoli@@ens-lyon.fr}
\end{center}

\bigskip 

\begin{abstract}
The issue of the number, nature and sequence of phase transitions in the
fully frustrated $XY$ (FFXY) model is a highly non trivial one due to the
complex interplay between its continuous and discrete degrees of freedom. In
this contribution we attack such a problem by means of a twisted conformal
field theory (CFT) approach \cite{cgm2} and show how it gives rise to the $%
U\left( 1\right) \otimes Z_{2}$ symmetry and to the whole spectrum of
excitations of the FFXY model \cite{a2}.
\end{abstract}

\newpage

\section{Introduction: the state of art}

The phase diagram of the FFXY model has been the subject of intensive
studies in the last thirty years, due to the presence of the mixed symmetry $%
U\left( 1\right) \otimes Z_{2}$. But a full and definitive answer to such a
question still lacks; in this paper we address the problem by means of a
twisted CFT approach \cite{cgm2}. The XY model on a square lattice in the
presence of an external magnetic field transversal to the lattice plane is
described by the action: 
\begin{equation}
H=-\frac{J}{kT}\sum_{\left\langle ij\right\rangle }\cos \left( \varphi
_{i}-\varphi _{j}-A_{ij}\right) ,  \label{act1}
\end{equation}
where $\left\{ \varphi \right\} $ are the phase variables on the sites, the
sum is over nearest neighbors, $J>0$ is the coupling constant and the bond
variables $A_{ij}=\frac{2e}{\hslash c}\int_{i}^{j}A{\cdot }dl$ satisfy the
full frustration condition $\sum_{plaquette}A_{ij}=\pi $. Choosing the
Landau gauge in such a way to get a lattice where each plaquette displays
one antiferromagnetic and three ferromagnetic bonds, we obtain two ground
states with opposite chiralities. The discrete $Z_{2}$ symmetry of the FFXY
model is broken at low temperature and will be restored beyond a certain
temperature after the formation of domain walls separating islands of
opposite chirality. The Ising transition overlaps to a vortex-unbinding
transition, which is associated with the continuous $U(1)$ symmetry \cite
{halsey2}. The action (\ref{act1}) can be rewritten as a fractionally
charged Coulomb gas (CG) defined on the dual lattice, $H_{CG}=-\frac{J}{kT}%
\sum_{r,r^{\prime }}\left( m\left( r\right) +\frac{1}{2}\right) G\left(
r,r^{\prime }\right) \left( m\left( r^{\prime }\right) +\frac{1}{2}\right) $%
, where $\lim_{\left| r-r^{\prime }\right| \rightarrow \infty }G\left(
r,r^{\prime }\right) =\log \left| r-r^{\prime }\right| +\frac{1}{2}\pi $ and
the neutrality condition $\sum_{r}\left( m\left( r\right) +\frac{1}{2}%
\right) =0$ holds. Such a model exhibits two possible phase transitions, an
Ising and a vortex-unbinding one. The issue whether there are two distinct
phase transitions, $T_{V}>T_{dw}$ or $T_{V}<T_{dw}$ with $T_{V}$ and $T_{dw}$
marking respectively the breaking of $U(1)$ and of $Z_{2}$ symmetry \cite
{double}, or a single transition with the simultaneous breaking of both
symmetries \cite{single} has been widely investigated. The model allows for
the existence of two topological excitations: vortices and domain walls.
Vortices are point-like defects such that the phase rotates by $\pm 2\pi $
in going around them \cite{halsey2}. A domain wall can be viewed as a line
on the square lattice, each segment of which separates two cells with the
same chirality \cite{korsh}. If the domain walls form a right angle, such
corners must behave as fractional vortices with topological charge $\pm 1/4$ 
\cite{halsey2}. There exists a temperature $T_{dw}$ such that, when $%
T>T_{dw} $, dissociation of bound pairs of fractional vortices is allowed,
which triggers the dissociation of pairs of ordinary vortices. The system
undergoes two phase transitions with temperatures such that $T_{V}<T_{dw}$ 
\cite{knops1}. At finite temperatures kinks may appear on the domain wall.
Simple kinks must behave as fractional vortices with topological charge $\pm
1/2$ while a double kink does not introduce mismatches in the phase
distribution. At low temperatures, all simple kinks are bound into neutral
pairs. As the temperature increases, a phase transition in the gas of
logarithmically interacting kinks leads to pair dissociation and emergence
of free simple kinks \cite{korsh}. That takes place at $T_{K}<T_{V}$ \cite
{olsson} and produces two distinct bulk transitions with $T_{V}<T_{dw}$. A
more general conclusion is reached by studying the coupled XY-Ising model 
\cite{granato,blote}, which is in the same universality class of the FFXY
model. Such a model can be introduced starting with a system of two XY
models coupled through a symmetry breaking term \cite{granato}: 
\begin{equation}
H=A\left[ \sum_{i=1,2}\sum_{\left\langle r,r^{^{\prime }}\right\rangle }\cos
\left( \varphi ^{(i)}(r)-\varphi ^{(i)}(r^{^{\prime }})\right) \right]
+h\sum_{r}\cos 2\left( \varphi ^{(1)}(r)-\varphi ^{(2)}(r)\right) .
\label{H-GK}
\end{equation}
The limit $h\rightarrow 0$ corresponds to a full decoupling of the fields $%
\varphi ^{(i)}$, $i=1,2$, while the $h\rightarrow \infty $ limit corresponds
to the phase locking $\varphi ^{(1)}(r)-\varphi ^{(2)}(r)=\pi j$, $j=1,2$;
as a consequence the model gains a symmetry $U(1)\otimes Z_{2}$ and its
Hamiltonian renormalizes towards the XY-Ising model \cite{granato}. Its
phase diagram \cite{granato,blote,baseilhac} is built up with three branches
which meet at a multi-critical point $P$. Two branches describe separate
Kosterlitz-Thouless (KT) and Ising transitions while the third ($PT$)
corresponds to single transitions with simultaneous loss of XY and Ising
order. It becomes a first order one at a tricritical point $T$ and seems to
be non-universal \cite{granato}; in fact the numerical estimate for the
central charge, $c\sim 1.60$, is higher than the value $c=3/2$, pertinent to
a superposition of critical Ising and gaussian models \cite{foda}. Indeed
the central charge seems to vary continuously from $c\approx 1.5$ near $P$
to $c\approx 2$ at $T$ \cite{blote}. The system lacks conformal invariance 
\cite{granato}, so one could consider suitable perturbations of the XY-Ising
model as a starting point to study the vicinity of the point $P$ \cite
{baseilhac}. Instead, recent numerical simulations on huge lattices \cite
{vicari} lead to two very close but separate transitions on the $PT$ line. A
possible solution could be \cite{knops1} the addition of an
antiferromagnetic coupling ($\overline{J}$) term to the Coulomb gas
Hamiltonian $H_{CG}$; for $\overline{J}\neq 0$, the two transitions on the $%
PT$ line separate with the KT one occurring at a lower temperature \cite
{knops1}. On the other side, by adding higher harmonics contributions to the
potential \cite{foda,sokalski}, the possibility of a merging critical point $%
T$ will be provided here in the context of a twisted CFT approach \cite{cgm2}%
, which extends the results of Ref. \cite{foda}, so recovering the whole
phase diagram \cite{granato,blote,baseilhac}.

\section{$m$-reduction procedure}

In this Section we recall those aspects of the twisted model (TM) which are
relevant for the FFXY model. We focus in particular on the $m$-reduction
procedure for the special $m=2$ case \cite{cgm2}, since we are interested in
a system with $U(1)\otimes Z_{2}$ symmetry. Such a theory describes well a
system consisting of two parallel layers of 2D electron gas in a strong
perpendicular magnetic field, with filling factor $\nu ^{(a)}=\frac{1}{2p+2}$
for each of the two $a=1$, $2$ layers \cite{cgm2}. Regarding the integer $p$%
, characterizing the flux attached to the particles, we choose the
``bosonic''\ value $p=0$, since it enables us to describe the highly
correlated system of vortices with flux quanta $\frac{hc}{2e}$. Let us start
from the ``filling'' $\nu =\frac{1}{2}$, described by a CFT with $c=1$ in
terms of a scalar chiral field $Q(z)=q-i\,p\,lnz+i\sum_{n\neq 0}\frac{a_{n}}{%
n}z^{-n}$, compactified on a circle with radius $R^{2}=1/\nu =2$; here $%
a_{n} $, $q$ and $p$ satisfy the commutation relations $\left[
a_{n},a_{n^{\prime }}\right] =n\delta _{n,n^{\prime }}$ and $\left[ q,p%
\right] =i$. From such a CFT (mother theory), using the $m$-reduction
procedure, which consists in considering the subalgebra generated only by
the modes in $Q(z)$ which are a multiple of an integer $m$, we get a $c=m$
orbifold CFT (the TM), which is symmetric under a discrete $Z_{m}$ group
and, for $m=2$, will be shown to describe the whole phase diagram of the
FFXY model. Its primary fields content, for the special $m=2$ case, can be
expressed in terms of two scalar fields given by: 
\begin{equation}
X(z)=\frac{1}{2}\left( Q(z)+Q(-z)\right) , \text{ } \phi (z)=\frac{1}{2}%
\left( Q(z)-Q(-z)\right) ;  \label{X}
\end{equation}
$X(z)$ is $Z_{2}$-invariant and describes the electrically ``charged''
sector of the new theory, while $\phi (z)$ satisfies the twisted boundary
conditions $\phi (e^{i\pi }z)=-\phi (z)$ and describes the ``neutral''
sector \cite{cgm2}. The TM primary fields are composite vertex operators $%
V\left( z\right) =U_{X}\left( z\right) \psi \left( z\right) $, where $%
U_{X}\left( z\right)$ is the vertex describing its ``charge''\ content and $%
\psi \left( z\right)$ describing the ``neutral''\ one. In the neutral sector
it is useful to introduce the two chiral operators $\psi \left( z\right)
\left( \overline{\psi }\left( z\right) \right)=\frac{1}{2\sqrt{z}}\left(
:e^{i\alpha \phi \left( z\right) }:\pm:e^{i\alpha \phi \left( -z\right)
}:\right)$, with only the first one obeying the boundary conditions. In a
fermionized theory they correspond to two $c=1/2$ Majorana fermions with
Ramond and Neveu-Schwartz boundary conditions \cite{cgm2} and, in the TM,
they appear to be not completely equivalent. In fact the whole TM theory
decomposes into a tensor product of two CFTs, a $Z_{2}$ invariant one with $%
c=3/2$ and symmetry $U(1)\otimes Z_{2}$ and the remaining $c=1/2$ one
realized by a Majorana fermion in the twisted sector. Furthermore the
energy-momentum tensor of the Ramond part of the neutral sector develops a
cosine term, $T_{\psi }\left( z\right) =-\frac{1}{4}\left( \partial \phi
\right) ^{2}-\frac{1}{16z^{2}}\cos \left( 2\sqrt{2}\phi \right)$, a
signature of a tunneling phenomenon which selects out the new stable $c=3/2$
vacuum; we identify such a theory with the one describing the FFXY model
conjectured in Ref. \cite{foda}.

\section{FFXY phase diagram from TM model}

In this Section we will derive the FFXY phase diagram in terms of the RG
flow which originates from perturbing our TM model with relevant operators.
We observe that the limit $h\rightarrow 0$ in the Hamiltonian of Eq. (\ref
{H-GK}) gives rise, in the continuum, to a CFT with two scalar boson fields $%
\varphi ^{(i)}$ and with central charge $c=2$. Now a good candidate to
describe the FFXY model at criticality around the point $T$ of the phase
diagram is a CFT, with $c=2$, which accounts for the full spectrum of
excitations of the model: vortices, domain walls, and kinks. The role of the
boundary conditions in the description of the excitation spectrum is
crucial. In fact, by imposing the coincidence between opposite sides of the
square lattice, we obtain a closed geometry, which is the discretized
analogue of a torus and gives rise, for the ground state, to two
topologically inequivalent configurations, one for even and the other one
for odd number of plaquettes. So the ground state on the square lattice maps
into the ground state for the even case while it generates two straight
domain walls along the two cycles of the torus for the odd case. Such a
behaviour has to be taken into account by non trivial boundary conditions on
the field $\varphi ^{(i)}$ at the borders of the finite lattice. To this
aim, let $(-L/2,0),$ $(L/2,0)$, $(L/2,L)$, $(-L/2,L)$ be the corners of the
square lattice ${\cal L}$ and assume that the fields $\varphi ^{(i)}$
satisfy the boundary conditions $\varphi ^{(1)}(r)=\varphi ^{(2)}(r)$ for $%
r\in {\cal L}\cap {\bf x}$, ${\bf x}$ being the $x$ axis. That allows us to
consider the two fields $\varphi ^{(1)}$ and $\varphi ^{(2)}$ on the square
lattice ${\cal L}$ as the folding of a single field ${\cal Q}$, defined on
the lattice ${\cal L}_{0}$ with corners $(-L/2,-L),$ $(L/2,-L)$, $(L/2,L)$, $%
(-L/2,L)$. We can implement a discrete version of the $2$-reduction
procedure by defining the fields ${\cal X}(r)=\frac{1}{2}\left( {\cal Q}(r)+%
{\cal Q}(-r)\right)$, $\Phi (r)=\frac{1}{2}\left( {\cal Q}(r)-{\cal Q}%
(-r)\right)$ ($r\in {\cal L}_{0}$), which are symmetric and antisymmetric
under the action of the group $Z_{2}$. The Hamiltonian (\ref{H-GK}) can be
rewritten in terms of these fields and, for $h=0$, it gives rise, in the
continuum, to the TM action ${\cal A}=\int \left[ \frac{1}{2}\left( \partial 
{\cal X}\right) ^{2}+\frac{1}{2}\left( \partial \Phi \right) ^{2}\right]
d^{2}x$. Let us now show how the phase diagram of the FFXY model can be
described by the action: 
\begin{equation}
{\cal A}=\int \left[ \frac{1}{2}\left( \partial {\cal X}\right) ^{2}+\frac{1%
}{2}\left( \partial \Phi \right) ^{2}+\mu \cos \left( \beta \Phi \right)
+\lambda \cos \left( \frac{\beta }{2}\Phi +\delta \right) \right] d^{2}x%
\text{,}  \label{X-two-SG}
\end{equation}
which embodies the higher harmonic potential term conjectured in Refs. \cite
{foda,sokalski,knops1}. We assume the constraints $\beta ^{2}<8\pi $, which
characterizes both the cosine terms as relevant perturbations, and $\left|
\delta \right| \leq \pi /2$ \cite{mussardo}. Thus the ``neutral''\ sector is
a two-frequency sine-Gordon theory that can be viewed as a deformation of a
pure sine-Gordon one with the perturbing term $\lambda \cos \left( \beta
\Phi /2+\delta \right) $. The ultraviolet (UV) fixed point $\mu =0,$ $%
\lambda =0$ of the action (\ref{X-two-SG}) corresponds to the TM model with
central charge $c=2$, describing the fixed point $T$ in the FFXY phase
diagram. In order to study the RG flow in the ``neutral''\ sector let us
define the dimensionless variable $\eta \equiv \lambda \mu ^{-\left( 8\pi
-(\beta /2)^{2}\right) /\left( 8\pi -\beta ^{2}\right) }$; when $\eta =0$
the ``neutral''\ sector reduces to a sine-Gordon model with a particle
spectrum built of solitons and antisolitons and, for $\beta ^{2}<4\pi $,
some breathers. Switching on the perturbation a confinement of solitons into
states with zero topological charge takes place and packets formed by $2$ of
the original solitons survive as stable excitations for generic values of $%
\left| \delta \right| <\pi /2$. In the limit $\eta \rightarrow \infty $ the
2-soliton evolves into the $1$-soliton of the pure sine-Gordon model with $%
\mu =0$. An unbinding phenomenon takes place in the case $\delta =\pm \pi /2$
for finite $\eta $ and the $2$-soliton decomposes into a sequence of two
kinks $K_{1}$. So the existence of an intermediate critical value $\eta
=\eta _{c}$ is required at which a phase transition takes place and the RG
flow ends into the infrared (IR) fixed point described by a CFT with central
charge $c=1/2$, the Ising model. The central charge of the full model (\ref
{X-two-SG}) so changes from $c=2$ of the UV fixed point to $c=3/2$ of the IR
fixed point, i.e. we recover early known Monte Carlo results \cite{blote}.
Such an IR fixed point coincides with the $U(1)\otimes Z_{2}$ symmetric
component of the TM model, which results then to properly describe \cite
{foda} the fixed point $P$ in the phase diagram of the FFXY model.



\end{document}